\documentstyle[12pt,epsfig,graphicx]{article}
\def\babar{\mbox{{\em B}\kern-0.07cm{\footnotesize\em A}\kern-0.07cm{\em B}\kern-0.07cm{\footnotesize\em A\kern-0.07cmR}} }
\begin{document}
\thispagestyle{empty}
\setcounter{page}0

~\vfill
\begin{center}
{\Large\bf  Can we use hadronic $\tau$-decay\\ for $V_{us}$
determination} \vfill

{\large Mihail V.~Chizhov}
 \vspace{1cm}

{\em Center of Space Research and Technologies,\\ Faculty of Physics,
University of Sofia,\\ 1164 Sofia, Bulgaria}

\end{center}  \vfill

\begin{abstract}

It is known that the discrepancy in pion spectral functions extracted
from $e^+e^-$ annihilation and $\tau$-decay leads to different
predictions for the muon anomalous magnetic moment. We will show
that this discrepancy effects also the extraction of the Cabibbo
angle from the hadronic
$\tau$-decays. The corrections to the $\tau$ branching fractions,
corresponding to the presence of new {\em centi-weak\/}
tensor interactions, allow us to extract the Cabibbo angle from
$\tau$ decays in agreement with its other precision determinations.
Thus a more precise value of $|V_{us}|=0.2246\pm 0.0012$ is obtained
and as a consequence
$f_+(0)=0.9645\pm 0.0055$ and $F_K/F_\pi=1.196\pm 0.007$.

\end{abstract}

\vfill

\newpage
\section{Introduction}

The vector transitions in weak decays play a special role in
phenomenology of the weak interactions. According to the Standard
Model (SM) the $W$-boson and the photon originate from the same
multiplet of $SU(2)_L$ group and they have the same form of vector
coupling with matter. Therefore, the vector weak transition could be
related to its isospin analogous
electromagnetic process by the CVC hypothesis~\cite{CVC}.

For example, the branching ratio of $\tau^-\to\rho^-\nu_\tau$ decay can be
predicted on the base of the pure electromagnetic process
$e^+e^-\to\gamma^*\to\rho^0$~\cite{Tsai}. Indeed, earlier,
at a moderate precision of the experimental data, the agreement between these
two processes was satisfactory~\cite{Ivanchenko}. However, with
increasing the precision in both sets of experimental data the discrepancy
between them has reached an unacceptable level of 4.6
$\sigma$~\cite{DEHZ1}.

At present, even after the appearing of additional high-precision data,
the same level of discrepancy still remains~\cite{Davier}.
The importance of this discrepancy increases due to new very
accurate measurements of the muon anomalous magnetic
moment~\cite{g-2}. The latter should be compared with the SM
prediction, which include the evaluation of the hadronic contribution to
$(g-2)_\mu$ based on $e^+e^-$ or/and $\tau$ data. While $\tau$-based
data predict a somewhat close to the experimentally measured $(g-2)_\mu$
value, $e^+e^-$ data differs by 3.3 $\sigma$ from it.

To find a solution to the discrepancy between the description of
$e^+e^-$ and $\tau$ processes, some authors~\cite{iso} have assumed
that the problem is mainly due to additional isospin breaking
effects and not to experimental ones. However, it is still
possible that both data sets could be plagued with unknown yet
systematic errors, therefore, others directly advocate $\tau$ data
as more reliable~\cite{Maltman1}. In this letter we assume that,
in spite of some inconsistency in both data sets among different
experiments, they properly describe real physical processes, and
that the discrepancy stems from additional new {\em centi-weak\/}
tensor interactions in the weak processes.

Of course, so radical solution of the problem should effect also
other low-energy precise experimental data as on muon and neutron
decays, and so on. It has been shown in previous
papers~\cite{muon,neutron} that those effects are still within
experimental uncertainties of the present measurements. Conspiracy of
the hypothetical interactions is stipulated by their different
chiral properties in comparison with $V-A$ interactions. In many
cases this prevents or suppresses interference between the new
interactions and the ordinary ones, which should give a leading
contribution from the new interaction in the case of its weakness.
The only known examples of manifestation of such interference are
the radiative pion decay~\cite{discovery}, $\pi\to e\nu_e\gamma$, and
the $\tau$-decay~\cite{tau}, $\tau\to\rho\nu_\tau$.

In this paper it will be shown that the anomaly in the latter decay
has direct relation to the problem of $V_{us}$ extraction from the
$\tau$-decay: its systematically low value in comparison with other
evaluations. The recent \babar\cite{BaBar} and Belle~\cite{Belle}
measurements of branching fractions for several hadronic
$\tau$-decay modes with kaons allow to emphasize this fact~\cite{Banerjee}.

Assuming the presence of the new interactions we will analyse the most
precise extractions of the Cabibbo angle from superallowed $0^+\to
0^+$ nuclear beta transitions, neutron and $\tau$ decays, and also
from the ratio of kaon and pion decay widths,
\mbox{$\Gamma(K\to\mu\nu)$}/$\Gamma(\pi\to\mu\nu)$, $K_{l3}$ and hyperon
decays. The latter three methods strongly depend on
flavor-$SU(3)$--breaking effects, which cannot be safety calculated
at present, and do not provide a unique absolute determination of the
Cabibbo angle. Nevertheless, using the unitarity relation we will show
that all mentioned methods can lead to self-consistent results.

\section{New interactions and direct $V_{ud}$ determination}

The effective tensor interactions
\begin{eqnarray}\label{udenu}
  {\cal L}^{\rm eff}_T=\hspace{-0.2cm}&-&\hspace{-0.2cm}\sqrt{2}f_T
  G_F\,\bar{u}_L\sigma_{ml}d_R\, \frac{4q^l q_n}{q^2}\,
  \bar{e}_R\sigma^{mn}\nu_L
  \nonumber\\
  &-&\hspace{-0.2cm}\sqrt{2}f_T G_F\,\bar{u}_R\sigma_{ml}d_L\,
  \frac{4q^l q_n}{q^2}~ \bar{e}_R\sigma^{mn}\nu_L+{\rm h.c.}
\end{eqnarray}
naturally appear in an extension of the SM with new type of spin-1
chiral bosons described by the antisymmetric second-rank tensor
fields~\cite{MPL}. Here $q_m$ is the momentum transfer between quark
and lepton currents. The dimensionless coupling constant $f_T$
determines the strength of the new interactions relative to the
ordinary weak interactions. With magnitude $f_T\simeq 10^{-2}$ these
interactions allow to explain the anomaly in the radiative pion
decay~\cite{Bolotov} and at the same time to escape constraints from
other precise measurements, due to their peculiar form. In order
to distinguish these interactions from the ordinary ones, 
taking into account their strength, we will refer to them as the {\em
centi-weak\/} interactions.

Assuming flavor universality we can rewrite the interactions
(\ref{udenu}) for the third lepton family as
\begin{equation}\label{udtaunu}
  {\cal L}^{\rm eff}_T=
  -\sqrt{2}f_T G_F\,\bar{u}\sigma_{ml}d\,
  \frac{4q^l q_n}{q^2}~ \bar{\tau}_R\sigma^{mn}\nu_L+{\rm h.c.}.
\end{equation}
Namely, these interactions are responsible for an additional
contribution to the vector transition $\tau^-\to\pi^-\pi^0\nu_\tau$
in comparison with the SM prediction, based on the CVC hypothesis.
In the ref.~\cite{tau} simple formula for the corresponding
branching fractions has been derived
\begin{equation}\label{excess}
    {\cal R}_{2\pi}\equiv\frac{{\cal B}^{\rm exp}_{2\pi}-{\cal B}^{\rm CVC}_{2\pi}}
    {{\cal B}^{\rm CVC}_{2\pi}}
    =F_T\,\frac{6}{2+r^2}+F^2_T\,\frac{1+2r^2}{(2+r^2)r^2},
\end{equation}
where $r=m_\tau/m_\rho$ is the mass ratio of the $\tau$ lepton and
the $\rho$ meson, and the positive factor
\begin{equation}\label{FT}
    F_T=4r f_T \frac{f^\perp_\rho}{f^\parallel_\rho}>0,
\end{equation}
is expressed through the effective coupling constant $f_T$ and the
ratio of the transverse and longitudinal coupling constants of the
$\rho$ meson,
$f^\perp_\rho/f^\parallel_\rho=0.703^{+0.004}_{-0.007}$~\cite{TNJL}.

The knowledge of the experimental excess,
${\cal R}_{2\pi}=(3.8\pm 0.9)\%$~\cite{Davier}, allows to fix
\begin{equation}\label{fT}
    f_T=(0.69\pm 0.16)\times 10^{-2}
\end{equation}
more precisely than before, which is now up to 4$\sigma$ above zero.

The interactions (\ref{udenu}) do not effect the superallowed Fermi
transitions, which are described by a vector quark current in the
effective weak lagrangian
\begin{equation}\label{effW}
  {\cal L}_W^{\rm eff}=-\sqrt{2}G_F\, V_{ud}~
  \bar{u}\gamma_m(1-\gamma^5)d~
  \bar{e}_L\gamma^m\nu_L+{\rm h.c.}
\end{equation}
Therefore, the extraction of the matrix element $V_{ud}$ from the
supperallowed $0^+\to 0^+$ nuclear $\beta$-decays should give, in
principle, its more reliable value. However, in reality the product
$G_F V_{ud}$ is measured, where $G_F$ is the well known Fermi coupling
constant. The latter is deduced from the muon decay, which may be
effected by new interactions.

Indeed, assuming again universality of the new interactions (\ref{udenu})
one can introduce purely leptonic tensor interactions
\begin{equation}\label{munuenu}
  {\cal L}^{\rm lepton}_T=-\sqrt{2}f_T
  G_F\,\bar{\nu}_L\sigma_{ml}\mu_R\, \frac{4q^l q_n}{q^2}\,
  \bar{e}_R\sigma^{mn}\nu_L+{\rm h.c.},
\end{equation}
which include only left-handed neutrinos and right-handed charged
leptons. In accordance with the see-saw mechanism, we assume here that
the right-handed neutrinos are very massive and they do not contribute
to the low-energy processes. Therefore, in contrast to the semileptonic
tensor interactions, the effective leptonic interactions do not
include a term analogous to the second term in (\ref{udenu}). Let us
note that the interactions (\ref{munuenu}) are not present among the
Michel local interactions and their effect on electron spectrum cannot
be described by the Michel parameters only.

Besides the distortion of the ordinary $V-A$ electron spectrum, the
tensor interactions give additional positive contribution to the
muon decay width~\cite{muon94}
\begin{equation}\label{Gmuon}
  \Gamma_\mu=\Gamma^{\rm SM}_\mu
  \left\{1+12\,\frac{m_e}{m_\mu}f_T+3 f^2_T\right\}.
\end{equation}
Since $\Gamma_\mu\propto G^2_F$ this leads effectively to new value
for the real magnitude of the Fermi coupling constant
\begin{eqnarray}\label{realGF}
  G_F&=&G_F^{\rm exp}/\sqrt{1+12\,\frac{m_e}{m_\mu}f_T+3f^2_T}
  \nonumber\\
  &=& G_F^{\rm exp}/(1.00027\pm 0.00008)
  =(1.16605\pm 0.00009)\times 10^{-5}~{\rm GeV}^{-2}
\end{eqnarray}
instead of $G_F^{\rm exp}=(1.16637\pm 0.00001)\times 10^{-5}~{\rm GeV}^{-2}$.
Therefore, the master formula for the
$V_{ud}$ determination from superallowed $\beta$ decays~\cite{MS}
should be corrected as
\begin{equation}\label{masterVud}
  \vert V_{ud}\vert_\beta=(1.00027\pm 0.00008)~
  \sqrt{\frac{(2984.48\pm 0.05)~{\rm s}}{ft(1+RC)}}.
\end{equation}

The present most precise PDG value of $|V_{ud}|^{\rm PDG}=0.97377\pm
0.00027$~\cite{PDG} is based on nine best measured superallowed
decays~\cite{beta9,Savard} without the mentioned correction.
However, there exist a problem with the recent precise Penning-trap
measurements~\cite{Savard,Eronen} of the $Q_{EC}$ value for the
superallowed decay of $^{46}$V, which has significantly effected its
$ft$ and $V_{ud}$ values. Recently, Towner and Hardy~\cite{TH} have
recalculated the isospin-symmetry-breaking correction including {\em
core orbitals}, which remove the discrepancy with $^{46}$V, and get
somewhat higher value for $|V_{ud}|^{\rm exp}_\beta=0.97418\pm
0.00026$.\footnote{The new calculations~\cite{TH} demonstrate a good
agreement among the corrected $ft$ values for the thirteen well
measured transitions, although there is a possible small discrepancy
for the cases of $^{50}$Mn and $^{54}$Co, which probably are
connected again to the older measurements of $Q_{EC}$ values.}
If we assume the latter value as a more reliable and apply new physics
correction from eq.~(\ref{masterVud}), this gives a new value of
\begin{equation}\label{newVud}
  \vert V_{ud}\vert_\beta=0.97444\pm 0.00026 \pm 0.00008_{f_T}=0.97444\pm 0.00027.
\end{equation}

Another independent information about the matrix element $|V_{ud}|$ can be
obtained from the neutron decay, which, however, besides the vector
transition, includes also the axial-vector amplitude characterized by
the additional phenomenological parameter $\lambda$. Therefore, besides the
precise knowledge of the neutron lifetime $\tau_n$ additional measurements
of $\lambda$ are necessary. The master formula for $|V_{ud}|$ determination
from the neutron decay reads~\cite{MS}
\begin{equation}\label{masterN}
    |V_{ud}|_n=\sqrt{\frac{(4908.7\pm 1.9)~{\rm s}}{\tau_n(1+3\lambda^2)}}
\end{equation}

At present there are severe disagreements in
the experimental results both for the neutron lifetime, $\tau_n$, and
$\lambda$ value and there is no sense to evaluate the averages.
According to us this situation is connected to huge corrections,
which have been applied in older experiments to the raw data in order to
extract the final result. By our opinion the recent measurements of
the neutron lifetime $\tau_n=878.5(8)$ s~\cite{Serebrov}
and parameter $\lambda^{\rm exp}=-1.2762(13)$~\cite{PERKEO2006}
are most reliable, where for the latter the applied corrections were just 0.09\%,
which is even below the experimental uncertainty. Using these numbers and
eq.~(\ref{masterN}) one gets
\begin{equation}\label{nVudSM}
    |V_{ud}|^{\rm SM}_n=0.97430\pm 0.00103.
\end{equation}
This value is in a very good agreement with (\ref{newVud}), while
the PDG value of the neutron lifetime $\tau^{\rm PDG}_n=885.7(8)$ s
leads to the following $|V_{ud}|_n=0.97033\pm 0.00103$, which is
about 4$\sigma$ below than the reference value
(\ref{newVud}).\footnote{The new value of the neutron lifetime is
favored also by cosmological reasons \cite{BBN}.}

Even the effect of the new tensor interactions cannot help to
improve the situation. It is known from ref.~\cite{neutron}
that the new interactions (\ref{udenu}) do not effect the neutron lifetime in the
leading order on $f_T$, while they change effectively
$\lambda=\lambda^{\rm exp}+\delta\lambda$, extracted from
the electron asymmetry parameter $A$,
where $\delta\lambda\simeq 0.0012$ is within
the experimental uncertainty for the reference value (\ref{fT}).
If we accept this correction, we get slightly different value than
(\ref{nVudSM})
\begin{equation}\label{neutronVud}
    |V_{ud}|_n=0.97506\pm 0.00103,
\end{equation}
which is also in agreement with (\ref{newVud}).

Using the most precise value of $|V_{ud}|$ (\ref{newVud})
and the unitarity relation
\begin{equation}\label{unitarity}
  \vert V_{ud}\vert^2+\vert V_{us}\vert^2+\vert V_{ub}\vert^2=1
\end{equation}
we can check selfconsistency of $V_{us}$ determination from
different processes. This is possible due to the small value of
$\vert V_{ub}\vert^{\rm PDG}=(4.31\pm 0.30)\times 10^{-3}$~\cite{PDG} and
present accuracy of $V_{ud}$ extraction (\ref{newVud}). Indeed,
neglecting $\vert V_{ub}\vert^2$ effects the unitarity relation
(\ref{unitarity}) at an accuracy level of $2\times 10^{-5}$, which is
much less than the uncertainty $2\vert V_{ud}\vert\delta\vert
V_{ud}\vert\simeq 53\times 10^{-5}$ stems from $\vert V_{ud}\vert^2$
contribution. Therefore, the approximation by the Cabibbo angle $\vert
V_{ud}\vert=\cos\theta_C$ and $\vert V_{us}\vert=\sin\theta_C$ is
still very suitable for description of processes with light
quarks.

\begin{figure}[th]
\epsfig{file=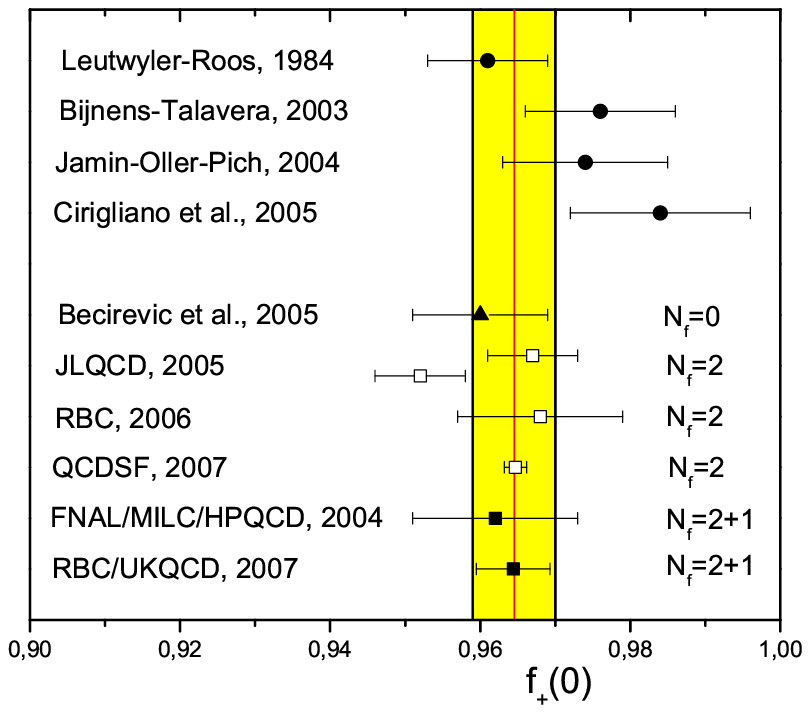,width=7.7cm} \epsfig{file=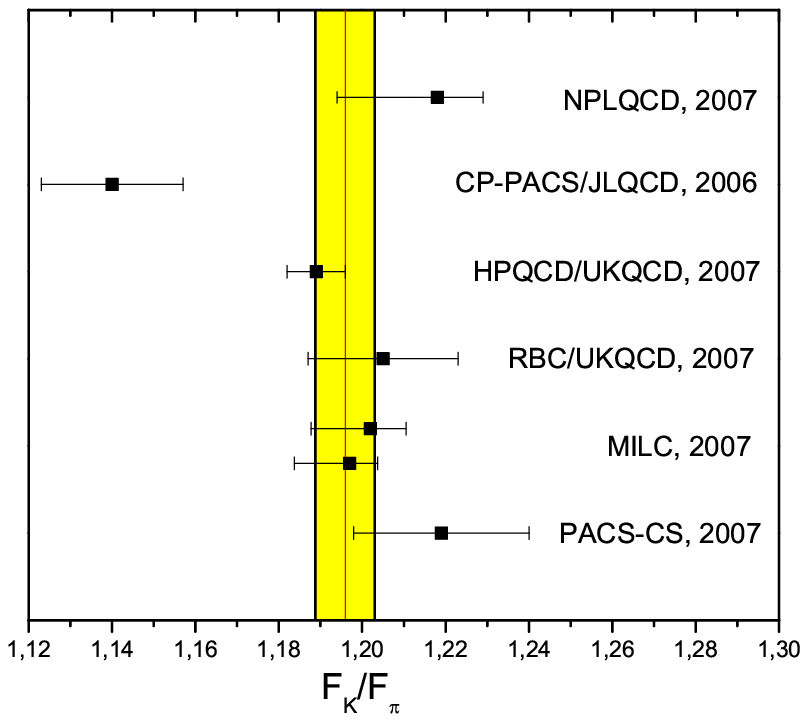,width=7.35cm}
\caption{\label{fig:1} The analytical~\cite{Roos,ChPT} and lattice~\cite{Becirevic+}
results for $f_+(0)$ (left) and the unquenched $N_f=2+1$ lattice calculations
of $F_K/F_\pi$ \cite{fKpi,JLQCD} (right) versus the corresponding predicted
values (\ref{f0}) and (\ref{fKpi}).}
\end{figure}

Using this approximation and the corrected $V_{ud}$ value (\ref{newVud}),
extracted from the superallowed $\beta$ decay, one gets matrix element
\begin{equation}\label{newVus}
  \vert V_{us}\vert^{\rm uni}_\beta
  =0.22463\pm 0.00112\pm 0.00033_{f_T}=0.22463\pm 0.00117,
\end{equation}
which is in agreement with the PDG value of $\vert
V_{us}\vert^{\rm PDG}=0.2257\pm 0.0021$, but has almost two times
better accuracy.
Since the most precise value of $\vert V_{us}\vert$
is extracted from semileptonic kaon decays $K_{e3}$, namely from the
expression $\vert V_{us} f_+(0)\vert = 0.21666\pm
0.00048$~\cite{Mescia}, our result (\ref{newVus}) indirectly
confirms the reliability of the analytical evaluations~\cite{Roos,ChPT}
and the lattice calculations~\cite{Becirevic+} of $SU(3)$-breaking
corrections to the form factor (see the left panel in the Fig.~\ref{fig:1})
\begin{equation}\label{f0}
    f_+(0)=0.9645\pm 0.0053\pm 0.0014_{f_T}=0.9645\pm 0.0055,
\end{equation}
which is in a very good agreement with the naive average of the results,
presented in the Fig.~\ref{fig:1}
\begin{equation}\label{avef0}
    f^{\rm ave}_+(0)=0.9646\pm 0.0013.
\end{equation}
Assuming the latter as an independent input, we can derive the
{\em most\/} precise values of the matrix elements
\begin{equation}\label{mostVus}
    \vert V_{us}\vert^K=0.22462\pm 0.00058
\end{equation}
and
\begin{equation}\label{mostVud}
    \vert V_{ud}\vert^K=0.97445\pm 0.00013
\end{equation}
from the kaon decays only.

Another verification of the results in eqs.~(\ref{newVud}) and
(\ref{newVus}), can be found from the experimental ratio of the
decay widths
$\Gamma(K^+\to\mu^+\nu_\mu(\gamma))=3.3716(99)\times 10^{-14}$~ MeV
and $\Gamma(\pi^+\to\mu^+\nu_\mu(\gamma))=2.5281(5)\times 10^{-14}$~MeV~\cite{PDG}
which is not effected by the tensor interactions and
is proportional to the ratios
\begin{equation}\label{VusVud}
    \tan\theta_C\equiv\frac{\vert V_{us}\vert}{\vert V_{ud}\vert}=0.23052\pm 0.00121\pm
    0.00036_{f_T}=0.23052\pm 0.00126
\end{equation}
and
\begin{equation}\label{fKpi}
    \frac{F_K}{F_\pi}=1.196\pm 0.007\pm 0.002_{f_T}=1.196\pm 0.007.
\end{equation}
For the latter ratio there are several unquenched lattice calculations
for $N_f=2+1$ dynamical quark flavors~\cite{fKpi}, which are in a good
accordance with eq.~(\ref{fKpi}), except the CP-PACS/JLQCD
result~\cite{JLQCD} (see the right panel in the Fig.~\ref{fig:1}).
Their naive average
\begin{equation}\label{aveFKPi}
    \frac{F_K}{F_\pi}\Bigg\vert^{\rm ave}=1.192\pm 0.005
\end{equation}
also agrees with eq.~(\ref{fKpi}).

It is interesting to note, that many authors of the lattice
calculations compare their results with the old `experimental' value
$(F_K/F_\pi)^{\rm exp}=1.223(12)$ from review \cite{Suzuki}, which is still
present in the recent PDG editions and is used as an important input
parameter for the chiral perturbation theory. This value is based on the
old result by Leutwyler and Roos for $\vert V_{us}\vert^{\rm
old}=0.2196(26)$~\cite{Roos} from $K_{e3}$ decays, which is pushed up
at present to a higher central value after the recent decided
experiments by BNL E865~\cite{BNL}, KTeV, NA48, ISTRA+ and KLOE
Collaborations.\footnote{See also Leutwyler's remark~\cite{Leutwyler}
about the present {\em experimental\/} value of $F_K/F_\pi$
and paper \cite{BP}.}
Therefore, the more precise and updated result from eq.~(\ref{fKpi})
should be used.


\section{Tau decays}

Let us turn now to the $\tau$ decays. Being a heavy lepton, $\tau$
possesses pure leptonic decays as well as hadronic decays, which to
a less extent are effected by the QCD complications than the
hadronic meson decays. At the same time many of the hadronic $\tau$
decays could be considered as cross-channels of the semileptonic
meson decays or just related through the CVC hypothesis to the
processes of the electromagnetic $e^+ e^-$ annihilation. All
this makes the $\tau$ decay investigations very sensitive to an
eventual manifestation of new physics.

Indeed at present there is a serious well known 4.5$\sigma$
discrepancy between the measured and the predicted two pion
branching ratio of the $\tau$ decay~\cite{Davier}. Neither SUSY nor
other known models can explain this anomaly. However, it has a natural
explanation~\cite{tau} in the extended standard model with the effective
tensor interactions (\ref{udtaunu}). If we assume the presence of such
interactions, it will be possible also to solve the problem with
systematically low central value of $V_{us}$ extracted from hadronic
$\tau$ decays~\cite{Gamiz}.

The idea is very simple and is based on the proper accounting for the
additional contributions from the new semileptonic interactions (\ref{udtaunu})
and the
analogous pure leptonic interactions (\ref{munuenu}) for the $\tau$ lepton.
These interactions could contribute both to the
leptonic and hadronic modes of the $\tau$ decays. It is know that
only six decay modes $\tau^-\to e^-\bar{\nu}_e\nu_\tau$, $\tau^-\to
\mu^-\bar{\nu}_\mu\nu_\tau$, $\tau^-\to\pi^-\nu_\tau$,
$\tau^-\to\pi^-\pi^0\nu_\tau$, $\tau^-\to\pi^-2\pi^0\nu_\tau$ and
$\tau^-\to\pi^-\pi^+\pi^-\nu_\tau$ account for 90\% of the decays.
Let us consider them in detail.

\begin{figure}[th]
\epsfig{file=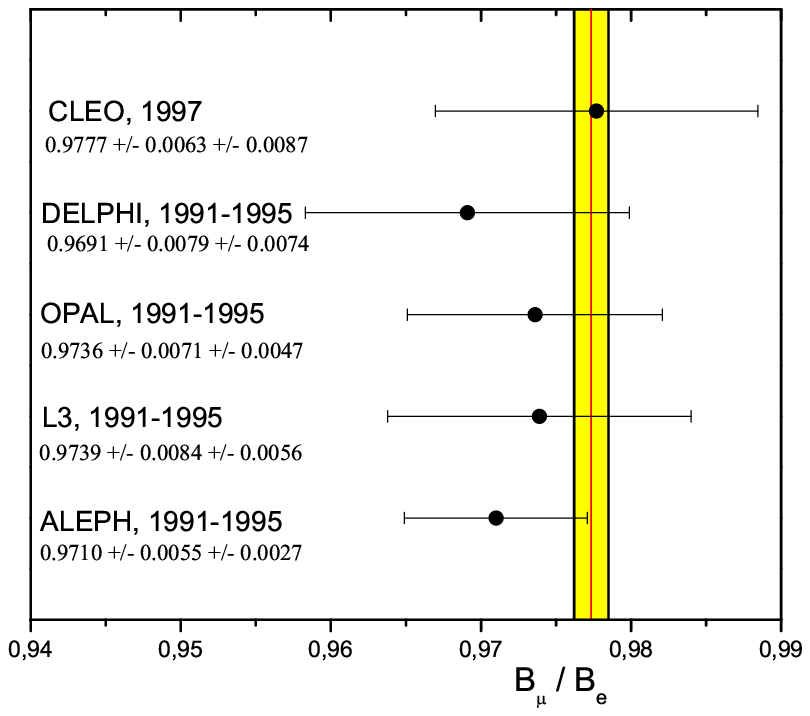,width=6.85cm}
\epsfig{file=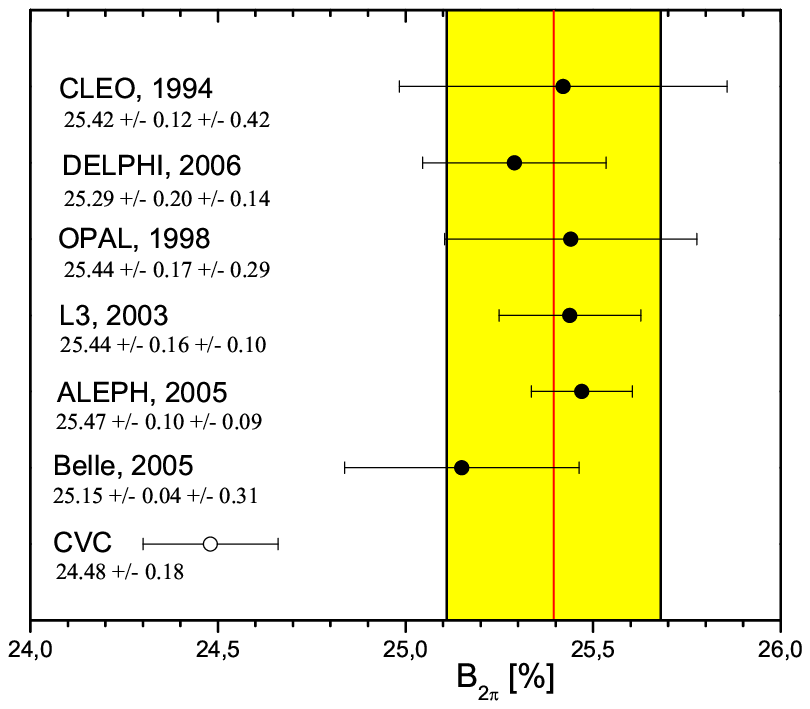,width=7.3cm} \caption{\label{fig:2} The
experimental ratios of the muonic and electronic decay widths
calculated from \cite{leptonBr} (left) and the two-pion branching
fractions from \cite{2piBr} (right) versus the predicted values.}
\end{figure}

According to the eq.~(\ref{Gmuon}) the leptonic decay widths of the
$\tau$ lepton should increase in comparison with the SM
due to new interactions by 0.02\%
and 0.51\% for the electronic and muonic modes, correspondingly.
This statement should be compared with the
experimental averages of the electronic and muonic branching fractions
\begin{equation}\label{expBeBm}
    {\cal B}_e^{\rm exp}=(17.821\pm 0.052)\%,\hspace{1cm}
    {\cal B}_\mu^{\rm exp}=(17.332\pm 0.049)\%,\hspace{1cm}
\end{equation}
which are claimed to be known with a relative precision of
0.3\%~\cite{DHZ} and their ratio ${\cal B}_\mu^{\rm exp}/{\cal
B}_e^{\rm exp}=0.9726\pm 0.0041$ perfectly matches the SM value
$0.972565(9)$. However, one should be cautions and mind
the {\em overconsistency\/} of leptonic branching fraction
measurements~\cite{hayes}. The predicted corrected ratio
\begin{equation}\label{rBmBe}
    \frac{{\cal B}_\mu}{{\cal B}_e}= 0.9774\pm 0.0011
\end{equation}
although higher than the experimental average, is still in agreement with
the data~(see the left panel in the Fig.~\ref{fig:2}).

All hadronic modes of $\tau$ decays into final state with {\em odd\/}
pions, which are related to the {\em axial-vector\/} hadronic currents,
are not effected by the new interactions due to the absence of an
appropriate tensor current $\bar{u}\sigma_{mn}\gamma^5 d$ in
(\ref{udtaunu}). At the same time all nonstrange ($\Delta S=0$) decays, which
undergo through hadron {\em vector\/} transitions, acquire additional
contributions. A substantial effect of the new interactions has been
found namely in $\tau^-\to\pi^-\pi^0\nu_\tau$ decay~\cite{tau} due
to the constructive interference between the new interactions and the
standard ones. The comparison of the experimental data with the corrected
CVC value using eq.~(\ref{excess})
\begin{equation}\label{corB2pi}
    {\cal B}^{\rm exp}=(25.40\pm 0.19\pm 0.22_{f_T})\%=(25.40\pm 0.29)\%
\end{equation}
is shown in the right panel of the Fig.~\ref{fig:2}.

The strange ($|\Delta S|=1$) hadronic $\tau$ decays, which contain around
3\% of the total decay width, are very important input for $V_{us}$
determination. They just concern this matrix element for Cabibbo-suppressed
(axial-)vector transitions between left-handed quarks of the different
generations. The new tensor interactions have different chiral structure than
the ordinary $V-A$ interactions and, in general, should be described by
a different mixing matrix between left-handed and right-handed quarks.
At present there is no indication of a presence of new interactions in kaon
decays~\cite{tensorK}. Therefore, we assume a near diagonal structure of
the new mixing matrix and an absence of tensor transitions between different
generations.

The master formula for the matrix elements extraction from the $\tau$ decays
has the following form
\begin{equation}\label{masterTau}
    \frac{R^{00}_{\tau,V+A}}{|V_{ud}|^2}-\frac{R^{00}_{\tau,S}}{|V_{us}|^2}
    \equiv\delta R_{\tau,\rm th},
\end{equation}
where $R^{00}_{\tau,V+A}$ is the ratio of the inclusive nonstrange hadronic
decay width and the electron width,
while $R^{00}_{\tau,S}$ is its strange counterpart.
It is assumed that only the ordinary $V-A$ interactions exist.
Here the difference $\delta R_{\tau,\rm th}$
is the $SU(3)$-breaking quantity induced mainly by the strange quark mass, which
can be theoretically estimated within the QCD framework~\cite{deltaR}.

The experimentally measured quantities $R_{\tau,V+A}$ and $R_{\tau,S}$ are not
independent and their sum should be related to the following combination of
the leptonic branching fractions
\begin{equation}\label{sumBr}
    R_\tau\equiv R_{\tau,V+A}+R_{\tau,S}=\frac{1-{\cal B}_e-{\cal B}_\mu}{{\cal B}_e}=
    3.635\pm 0.010,
\end{equation}
where the average electronic branching fraction
${\cal B}_e^{\rm ave}=(17.818\pm 0.032)\%$ \cite{DHZ} and
the predicted ratio (\ref{rBmBe}) were used.
The latter corrects the right hand side
of eq.~(\ref{sumBr}) for an effect of the tensor interactions
in the lepton channels, however their
impact is well below the uncertainty and slightly
changes only the cental value.

The left hand side of eq.~(\ref{sumBr}) also should be corrected for the additional
contributions from the new physics. Since the vector transitions occur through
the vector meson resonances $\rho$, which are effected by the new interactions,
then
not only $\tau^-\to\pi^-\pi^0\nu_\tau$ channel, but $\tau^-\to\pi^-3\pi^0\nu_\tau$,
$\tau^-\to2\pi^-\pi^+\pi^0\nu_\tau$ and other channels with {\em even\/} pions states
should be revised, as well. Using the recent evaluations~\cite{Davier}
\begin{equation}\label{BtauCVC}
    {\cal B}_\tau-{\cal B}_{e^+e^-}^{\rm CVC}=
    \left\{\begin{array}{ll}
             (0.92\pm 0.21)\% & {\rm for}~\tau^-\to\pi^-\pi^0\nu_\tau \\
             -(0.08\pm 0.11)\% & {\rm for}~\tau^-\to\pi^-3\pi^0\nu_\tau \\
             (0.91\pm 0.25)\% & {\rm for}~\tau^-\to2\pi^-\pi^+\pi^0\nu_\tau
           \end{array}
    \right.
\end{equation}
and neglecting the higher number meson states we get 5$\sigma$ deviation in
\begin{equation}\label{RtauCVC}
    R_{\tau,V}-R_{\tau,V}^{\rm CVC}=0.098\pm 0.019.
\end{equation}

If we assume that the undistorted vector contribution
in the $\tau$ decay can be deduced
from the experimental data on $e^+e^-$ annihilation by the CVC hypothesis,
i.\ e.\
$R^{00}_{\tau,V}=R^{\rm CVC}_{\tau,V}$, while its axial-vector and strange
counterparts are not effected by the new interactions, i.\ e.\
$R^{00}_{\tau,A}=R_{\tau,A}$ and $R^{00}_{\tau,S}=R_{\tau,S}$ respectively,
then we can apply corrections (\ref{RtauCVC}) and (\ref{sumBr}) to the master
eq.~(\ref{masterTau}) with
\begin{equation}\label{R00NS}
   R^{00}_{\tau,V+A}\equiv R^{00}_\tau-R^{00}_{\tau,S}=3.537(22)-R_{\tau,S}.
\end{equation}
The master equation is usually solved for the unknown parameter $|V_{us}|$,
while the
other entries are considered known inputs. Here we will
show that using the unitarity relation, it is possible to extract the ratio
\begin{eqnarray}\label{masterVusVud}
    \frac{|V_{us}|^2}{|V_{ud}|^2}&=&
    \frac{\sqrt{\left(R^{00}_\tau-\delta R_{\tau,\rm th}\right)^2
    +4 R^{00}_{\tau,S}R_{\tau,\rm th}}
    -\left(R^{00}_{\tau,V+A}-R^{00}_{\tau,S}
    -\delta R_{\tau,\rm th}\right)}{2 R^{00}_{\tau,V+A}}
    \nonumber\\
    &\approx&\frac{R^{00}_{\tau,S}}{R^{00}_{\tau,V+A}-
    \left(\frac{\textstyle R^{00}_{\tau,V+A}}{\textstyle R^{00}_\tau}\right)
    \delta R_{\tau,\rm th}}
\end{eqnarray}
from eq.~(\ref{masterTau}) and the matrix element
$|V_{us}|$ with even better precision.

\begin{figure}[th]
\hspace{3cm}\epsfig{file=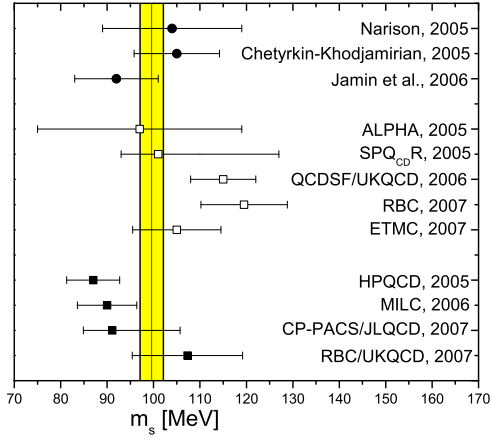,width=8cm} \caption{\label{fig:3}
The recent analytical \cite{Narison+} and lattice $N_f=2$ \cite{ALPHA+}
and $N_f=2+1$ \cite{HPQCD+} evaluations of $m_s(2~{\rm GeV})$.}
\end{figure}

To finalize the matrix elements extraction we need to define two more inputs
$R^{00}_{\tau,S}$ and $\delta R_{\tau,\rm th}$.
The latter depends on the strange quark
mass $m_s$. The recent $m_s$ evaluations, disregarding the
$\tau$ decays because they are
most effected by the new interactions,
are shown in the Fig.~\ref{fig:3}, which lead to the following
average mass
\begin{equation}\label{ms}
    m_s(2~{\rm GeV})=(99.6\pm 3.3)~{\rm MeV},
\end{equation}
where the uncertainty is corrected by the scale factor $S=1.3$,
and its corresponding
\begin{equation}\label{Rth}
    \delta R_{\tau,\rm th}=0.1602\pm 0.0046+(6.08\pm 1.00)\frac{m^2_s}{{\rm GeV}^2}
    =0.221\pm 0.012.
\end{equation}

The most critical part is the experimental value of $R_{\tau,S}$, which
after LEP epoch was
\begin{equation}\label{RsLEP}
    R^{\rm LEP}_{\tau,S}=0.1686\pm 0.0047,
\end{equation}
and at present after the new \babar\cite{BaBar} and Belle~\cite{Belle}
measurements is
\begin{equation}\label{newRs}
    R^{\rm new}_{\tau,S}=0.1617\pm 0.0040.
\end{equation}
Of course, these two values lead to different magnitudes for the ratio
\begin{equation}\label{tauVusVud}
    \frac{|V_{us}|}{|V_{ud}|}\Bigg\vert_\tau=\left\{
    \begin{array}{ll}
      0.2310\pm 0.0032 & {\rm from~eq.~}(\ref{RsLEP}) \\
      0.2260\pm 0.0028 & {\rm from~eq.~}(\ref{newRs})
    \end{array}
    \right.
\end{equation}
and for the matrix elements
\begin{equation}\label{tauVus}
    |V_{us}|_\tau=\left\{
    \begin{array}{ll}
      0.2251\pm 0.0029 & {\rm from~eq.~}(\ref{RsLEP}) \\
      0.2205\pm 0.0026 & {\rm from~eq.~}(\ref{newRs})
    \end{array}
    \right.
\end{equation}
and
\begin{equation}\label{tauVud}
    |V_{ud}|_\tau=\left\{
    \begin{array}{ll}
      0.97434\pm 0.00068 & {\rm from~eq.~}(\ref{RsLEP}) \\
      0.97540\pm 0.00058 & {\rm from~eq.~}(\ref{newRs})
    \end{array}
    \right. .
\end{equation}

Comparison with the eqs.~(\ref{newVud}), (\ref{newVus}) and (\ref{VusVud})
shows, that the extracted from the hadronic $\tau$ decays matrix elements,
based on the old data,
are in a perfect agreement with the present most precise measurements
from the superallowed beta decays, while the new \babar and Belle results
lead to the low value for $|V_{us}|$, but still statistically acceptable.
This agreement induces us to make doubt of the \babar and Belle measurements,
moreover, they have analyzed only several Cabibbo-suppressed
hadronic $\tau$ decay modes and
all their results show systematically low values.
Perhaps we need to wait for more serious investigations
which include thorough analysis of the main $\tau$ decay
channels as well.

Let us compare now our derivation (\ref{masterVusVud}) of the matrix elements
from the hadronic $\tau$ decays with the usually used one \cite{tauPRL}.
First of all let us note that the eq.~(\ref{masterVusVud}) for $|V_{us}|/|V_{ud}|$
determination is very similar to the formula for $|V_{us}|$ and even has a slightly
lower uncertainty. However, we get the main gain in precision for $|V_{us}|$
determination from the ratio $|V_{us}|/|V_{ud}|$ using unitarity.
The simple evaluations show that, if the precisions for $|V_{us}|$ and
$\tan\theta_C\equiv|V_{us}|/|V_{ud}|$ are the same, we get around 8\%
better accuracy for $|V_{us}|$ determination from the ratio
($\delta|V_{us}|=\delta\tan\theta_C/(1+\tan^2\theta_C)^{3/2}$).
The second advantage of
the eq.~(\ref{masterVusVud}) consists in a possibility of a
simultaneous extraction of $|V_{us}|$ and $|V_{ud}|$ as well with a good
precision. Although it cannot still compete with the accuracy
of the superallowed beta decays, we get $|V_{ud}|$ (\ref{tauVud})
with even less uncertainty than from the neutron decays (\ref{neutronVud}).

The surprising agreement between the matrix elements, extracted
from the superallowed beta decays and from the hadronic $\tau$ decays,
results from the essential corrections of the hadronic decay width, $R_\tau$,
(\ref{sumBr}) and (\ref{RtauCVC}) due to the new tensor interactions.
It should be noted that uncorrected
$R_\tau$ input leads approximately to one standard deviation less
central value for the
$|V_{us}|$, what has been pointed out in \cite{Maltman2}.
To support our
suggestions we will draw attention to the results of ref.~\cite{Maltman3},
where the weighted spectral integrals have been used. It has been shown
that using spectral weights which suppress the contributions from the region
of the spectrum with high hadronic masses, the central values of
the extracted $|V_{us}|$ are systematically higher than for the unweighted case.
It is just in concordance with our results from \cite{tau}, which
predict a monotonically increasing deviation of the $\tau$ spectral
function from the SM case with the increase of the hadron invariant mass.
Therefore, a suppression of this part of the spectrum allows to make more
reliable predictions.

In conclusion, we would like also to compare our results with $|V_{us}|$
extraction from strangeness-changing semileptonic hyperon decays,
which, as we assumed, are not influenced by the new interactions in contrast
with Cabibbo-allowed decays.
A simple phenomenological fit of the decay rates and axial-vector to vector
form factors ratios, up to $SU(3)$ breaking effects,
leads to the following result~\cite{CSW}
\begin{equation}\label{hypVus}
    |\tilde{f}_1 V_{us}|_{SU(3)}=0.2250\pm 0.0027,
\end{equation}
where $\tilde{f}_1$ is the ratio of the vector formfactor $f_1(0)$ to its
$SU(3)$ predicted value. This analysis has been confirmed in ref.~\cite{Mateu}.
However, the authors suggest two times bigger uncertainty due to the disagreement
of the different theoretical estimations of $\tilde{f}_1$.
Comparing the result (\ref{hypVus}) with the reference value (\ref{newVus})
one can conclude, that
$\tilde{f}_1$ is equal to one up to second-order $SU(3)$ breaking effects
in agreement with the Ademollo--Gatto theorem~\cite{AG}. It means that
the $SU(3)$ breaking corrections in hyperon decays at present accuracy
are still not essential and their account results in a smaller central value
of $|V_{us}|=0.2199\pm 0.0026$~\cite{Flores}.


\section{Conclusions}

The main purpose of this paper is to point out the necessity
for the account of the new tensor interactions in the weak processes.
For many processes their effect is still well below the
experimental uncertainties, however, for some of them their
account is urgently needed. It happens due to the weakness
of the new interactions and their different chiral
structure in comparison with the SM.
They can reveal themselves, for example, in chirally suppressed
pion decays, or in result of the interference of the
tensor currents from the {\em centi-weak\/} interactions with transverse
hadronic currents, as in tau decays. At the same time, due to
the complexity of the Lorentz structure of such interactions,
they do not contribute to the two-particle pion decay as the more
simple hypothetical pseudoscalar interaction would.
All these properties of the new tensor interactions allow
them often to escape stringent experimental constraints.

In this paper we have analyzed of the extraction of
the matrix elements between light quark species.
Using the available experimental data and the unitarity relation,
we have shown that all results can be come selfconsistent,
if

(a) the new data for the neutron decay are used,

(b) the old branching fractions of strangeness-changing $\tau$ decays
are correct,

(c) the additional tensor interactions are taken into account.\\
The latter removes the disagreement between the spectral functions extracted from
the electromagnetic $e^+e^-$ data and the $\tau$ decays. It gives us the possibility
to use both data for the evaluation of the hadronic contribution
into the muon anomalous magnetic moment. This could strengthen the present
discrepancy between its predicted and experimentally measured values.

Taking this discrepancy seriously we can constrain other sector of
our model connected with new {\em neutral\/} tensor currents, which should
also be present in the extension of the SM by doublets of the tensor
particles~\cite{MPL}. The new charged and neutral tensor current interactions
appear as effective interactions resulting from the exchanges of the massive
charged and neutral tensor particles, respectively.
The doublet structure of the new particles stems from the symmetry of the SM
and that the tensor interactions as the (pseudo)scalar ones mix
the left and right handed fermions from different representations
of $SU(2)_L\times U(1)_Y$ group. In order to compensate the contributions
of the new couplings of the tensor particles into the chiral anomaly,
the doubling of the doublets both for the tensor and Higgs particles is needed.
Besides this, in order to prevent also the presence of flavor-changing neutral currents,
up and down type fermions should couple different doublets of the Higgs~\cite{GW}
and tensor particles.

In ref.~\cite{masses} the masses of the new tensor particles have
been estimated. Two mass scales around 700 GeV and 1 TeV
for the different doublets have been derived on the basis of the dynamical
principe and the value of the effective coupling constant $f_T$, which shows
the relative strength of the new interactions in comparison with the SM.
Due to additional mixing the lightest charged tensor boson has
a mass around 500 GeV and can effect nonnegligibly the charged
current SM processes, while the lightest neutral tensor boson
couples just to the up type fermions and can effect
leptonic processes with neutrinos. Additional contributions
into neutral current processes with charged leptons
come from the exchanges of the heavy neutral tensor particles and are
around four times weaker than in the charged tensor current sector.

Nevertheless, an eventual mixing of the heavy neutral tensor particles
with photons can effect electromagnetic characteristics of
charged leptons like their anomalous magnetic moments, which have been
measured with unprecedent precision for the electron and muon.
It has been shown \cite{mixing}
that such mixing can explain the sign of the discrepancy between
the predicted and measured muon anomalous magnetic moment. In contrast
to the new heavy particles, which contribute through the loops and effect
the anomalous magnetic moments as $m^2_\ell/M^2_H$, the effect of
the mixing is proportional to $m_\ell/M_H$. Therefore, the effect
of the new physics on the electron anomalous magnetic moment could be
stronger than have been expected and can be tested in
independent measurements of the fine structure constant $\alpha$.
On the other hand the mass scale of the new tensor particles
offers a unique possibility for their discovery at the Large Hadron
Collider.

\section*{Acknowledgements}

I am grateful to K. E. Arms, J. L. Bijnens, I. R. Boyko, S. I. Eidelman,
S. Heinemeyer and Z. Zhang
for useful communications.

\pagebreak[3]

\end{document}